\documentclass[prd,superscriptaddress,twocolumn,showpacs,amsmath,%
preprintnumbers,showkeys]{revtex4-1}
\usepackage{bm} 
\usepackage{graphicx}
\usepackage{amssymb}
\usepackage{url}
\newcommand{\be}{\begin{equation}}
\newcommand{\ee}{\end{equation}}
\newcommand{\bea}{\begin{eqnarray}}
\newcommand{\eea}{\end{eqnarray}}
\begin{document}                                           
\title{Transverse nucleon structure and diagnostics of
hard parton--parton processes at LHC}
\author{L.~Frankfurt}
\affiliation{School of Physics and Astronomy, Tel Aviv University, 
Tel Aviv, Israel}
\author{M.~Strikman}
\affiliation{Department of Physics, Pennsylvania State University,
University Park, PA 16802, USA}
\author{C.~Weiss}
\affiliation{Theory Center, Jefferson Lab, Newport News, VA 23606, USA}
\date{September 13, 2010}
\begin{abstract}
We propose a new method to determine at what transverse momenta
particle production in high--energy $pp$ collisions 
is governed by hard parton--parton processes. Using information 
on the transverse spatial distribution of partons obtained from
hard exclusive processes in $ep/\gamma p$ scattering, 
we evaluate the impact parameter distribution of $pp$ collisions with a 
hard parton--parton process as a function of $p_T$ of the
produced parton (jet). We find that the average $pp$ impact parameters 
in such events depend very weakly on $p_T$ in the range 
$2 < p_T < \text{few 100 GeV}$, while they are much smaller 
than those in minimum--bias inelastic collisions. The impact 
parameters in turn govern the observable transverse multiplicity 
in such events (in the direction perpendicular to the trigger particle 
or jet). Measuring the transverse multiplicity as a function of $p_T$ 
thus provides an effective tool for determining the minimum $p_T$
for which a given trigger particle originates from a hard 
parton--parton process. Additional tests of the proposed
geometric correlations are possible by measuring the dependence
on the trigger rapidity. Various strategies for implementing 
this method are outlined.
\end{abstract}
\preprint{JLAB-THY-10-1228}
\maketitle
\section{Introduction}
The first experimental results from LHC once again raise the question
at what transverse momenta particle production in $pp$ collisions is
dominated by hard parton--parton interactions. A quantitative 
understanding of the relevant mechanisms is important not only 
for future studies of QCD phenomena, but also for controlling the 
strong interaction background in new particle searches. 
The challenge lies in the fact that the
growth of the average multiplicities makes it very difficult to
observe jets with moderate $p_T$, while at the same time the
properties of non--perturbative semi--hard dynamics and its ability to
produce particles with $p_T \sim$ few GeV are not well understood.

\begin{figure}[b]
\includegraphics[width=0.42\textwidth]{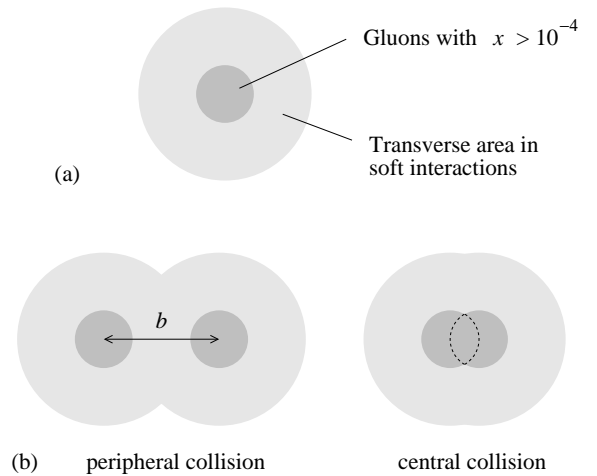}
\caption{(a) The two--scale picture of transverse nucleon structure at
high energies (transverse view). 
(b) Its implication for $pp$ collisions.  Peripheral
collisions constitute the dominant part of the overall inelastic cross
section. Hard processes happen predominantly in central collisions,
where the areas occupied by large--$x$ partons overlap.}
\label{fig:percent}
\end{figure}

In an earlier article \cite{Frankfurt:2003td}, we demonstrated that 
the nucleon's transverse partonic structure plays an essential role
in the theoretical analysis of $pp$ collisions with hard processes. 
Experiments in hard exclusive electroproduction of vector mesons 
$\gamma^\ast p \rightarrow V + p$ and photoproduction of heavy 
quarkonia $\gamma p \rightarrow J/\psi + p$ have shown that the 
gluons with $10^{-4} < x < 10^{-1}$ are localized at small transverse 
distances of $0.4-0.5 \, \text{fm}$ (median, depending 
on $x$ and $Q^2$), much smaller than the characteristic range of soft
interactions at high energies, see Fig.~\ref{fig:percent}a. Qualitatively, 
this is explained by the fact that Gribov diffusion in the partonic wave
function, which causes the range of soft interactions to grow with
energy \cite{Gribov:1973jg}, is suppressed for highly virtual
constituents. In $pp$ scattering this two--scale picture implies that
hard processes mostly occur in central collisions,
where the areas occupied by partons in the relevant $x$--range overlap.
Peripheral collisions constitute the dominant part of the overall 
inelastic cross section without contributing much to 
inclusive jet production, see Fig.~\ref{fig:percent}b. 
A trigger on a hard process thus, on average,
selects central $pp$ collisions \cite{Frankfurt:2003td}. 
Numerical studies show that at a center--of--mass energy 
$\surd s = 14 \, \textrm{TeV}$ a dijet trigger on $p_T \sim
\textrm{few 10 GeV}$ reduces the median $pp$ impact parameter
$b$ by a factor of $\sim 2$ compared to minimum--bias 
inelastic collisions; the reduction is nearly as strong at the 
current LHC energy of $7 \, \textrm{TeV}$ (see below).
%
%

Here we point out that these insights into the transverse geometry 
of $pp$ collisions can be used to address the question at what 
transverse momenta particle production is governed by hard 
parton--parton processes. The key observation is that the
transverse multiplicity, measured in the direction perpendicular 
to the transverse momentum of the trigger particle (or jet), 
is correlated with the average impact parameter in the underlying
$pp$ event. If the trigger particle originates from a hard
parton--parton process, the underlying $pp$ collision is central,
and the average impact parameter depends only weakly on $p_T$.
The transverse multiplicity thus should be practically 
independent of $p_T$, and substantially larger than that in
minimum bias events. If, however, the trigger particle is produced 
by soft interactions, the transverse multiplicity should  
be substantially smaller and reflect the average multiplicity in 
minimum--bias inelastic collisions. In this sense, the transverse
multiplicity can serve as a diagnostic of the dynamics in the
production of the trigger particle at given $p_T$.
First theoretical suggestions for studies of the transverse multiplicity
were put forward in Ref.~\cite{Marchesini:1988hj}. 
Experimental investigation of the correlation between the jet production 
and the structure of the underlying event were pioneered by 
the CDF experiment at the Tevatron \cite{Affolder:2001xt}.  
The first data on underlying event structure in collisions with 
hard processes at the LHC were recently reported by ATLAS \cite{EmilyNurse} 
and CMS \cite{Khachatryan:2010pv,Lucaroni}.

Additional tests of the geometric correlations described here
become possible with measurements of the dependence of the transverse
multiplicity on the rapidity of the trigger jets. In particular,
we predict that in the rapidity region not affected by the 
fragmentation of the trigger jets the enhancement of the
multiplicity will persist and be isotropic in transverse space. 
In this way one could verify the universality of particle production 
in the central $pp$ collisions selected by hard processes.

This article is organized as follows. In Sec.~\ref{sec:parton} we 
summarize our knowledge of the nucleon's transverse partonic structure
and update our parametrization of the transverse gluonic size as a 
function of $x$. In Sec.~\ref{sec:impact} we use this information to study 
the impact parameter distributions of $pp$ events with hard processes
in dependence on the transverse momentum $p_T$ of the trigger 
particle (jet), for the kinematics currently covered 
by LHC, and find that the median $b$ weakly depends on the trigger
$p_T$ and is substantially smaller than that in minimum--bias
inelastic collisions. In Sec.~\ref{sec:multiplicity} we discuss
the connection between centrality and the transverse multiplicity,
and how measurement of the latter provides an effective means
of quantifying at what $p_T$ particle production
is dominated by hard QCD processes. In Sec.~\ref{sec:rapidity}
we consider the dependence of the multiplicity on the rapidity
of the trigger, and how it can be used for additional tests
of the dominance of small impact parameters in $pp$ collisions
with hard processes. In Sec.~\ref{sec:future} we present several 
suggestions for further analysis of the $pp$ event structure data. 
A summary and discussion of our results are given in Sec.~\ref{sec:summary}.

Our analysis relies on information on the nucleon's transverse 
partonic structure obtained from hard exclusive processes
in $ep/\gamma p$ scattering. Extending our previous 
study \cite{Frankfurt:2003td}, we present here an updated 
parametrization of the transverse distribution of gluons, 
which takes into account the more recent HERA data 
\cite{Chekanov:2004mw,Aktas:2005xu} and permits realistic 
uncertainty estimates. The numerical results are nevertheless
close to those obtained in our previous study.

Current Monte Carlo (MC) generators for $pp$ events 
usually do not take into account the available experimental information on 
transverse nucleon structure and treat the distribution of gluons 
over transverse position as a free function.
The typical setting for PYTHIA \cite{PYTHIA} and HERWIG \cite{HERWIG}
correspond to a transverse area occupied by gluons which is a factor 
$\sim 2$ smaller than what is indicated by the HERA data (see below). 
In the analysis of experimental data, the shape of the transverse 
gluon distribution is usually treated as one of the tuning parameters, 
see e.g.\ Refs.~\cite{Affolder:2001xt,Khachatryan:2010pv}.
While we do not directly address these technical issues here, our results 
certainly have implications for the design of future MC
generators for $pp$ events at LHC.

\section{Transverse partonic structure of nucleon}
\label{sec:parton}
Information on the transverse spatial distribution of gluons in the
nucleon comes from the study of hard exclusive processes such as 
electroproduction of vector mesons, $\gamma^\ast p \rightarrow V + p$,
or the photoproduction of heavy quarkonia, $\gamma p \rightarrow 
J/\psi + p$. Thanks to a QCD factorization theorem \cite{Collins:1996fb}, 
the amplitude of these processes in the leading--twist approximation 
can be expressed in terms of the gluon generalized parton 
distribution (or GPD), which parametrizes
the matrix element for the emission and absorption of a gluon by
the target. Of particular interest is its $t$--dependence in the
``diagonal'' case of equal momentum fraction of the emitted and
absorbed gluons. It is described by the normalized 
two--gluon form factor $F_g (x, t| Q^2)$, where $x$ is the gluon 
momentum fraction and $t = -\bm{\Delta}_\perp^2$ 
the transverse momentum transfer to the target. This function can be
regarded as the transverse form factor of gluons with longitudinal 
momentum fraction $x$ in the nucleon. Its Fourier transform 
describes the transverse spatial distribution of gluons with 
given $x$,
\be
F_g (x, \rho | Q^2) \; \equiv \; \int\!\frac{d^2 \Delta_\perp}{(2 \pi)^2}
\; e^{i (\bm{\Delta}_\perp \bm{\rho})}
\; F_g (x, t = -\bm{\Delta}_\perp^2 | Q^2) ,
\label{rhoprof_def}
\ee
where $\rho \equiv |\bm{\rho}|$ measures the distance from the transverse
center--of--momentum of the nucleon, and the distribution is normalized
such that $\int d^2\rho \, F_g (x, \rho | Q^2) = 1$.

Experiments in hard exclusive processes actually probe the gluon
GPD in the non--diagonal case (different momentum fraction of the
emitted and absorbed gluon), because of the longitudinal momentum
transfer required by kinematics. At $x \lesssim 10^{-2}$, QCD 
evolution allows one to relate the non-diagonal gluon GPD to the 
diagonal one at the input scale, and the diagonal two--gluon form factor 
can be directly inferred from the $t$--dependence of the measured cross
sections. At larger $x$, the relation between the non--diagonal and
diagonal GPDs generally becomes model--dependent, but useful 
information can still be extracted using GPD parametrizations.

The $t$--dependence of the measured differential cross sections of
exclusive processes at $|t| < 1 \, \text{GeV}^2$ is commonly
described either by an exponential, or by a dipole form inspired 
by analogy with the nucleon elastic form factors. Correspondingly,
we consider here two parametrizations of the two--gluon form factor:
\be
F_g (x, t|Q^2) \;\; = \;\; 
\left\{ \begin{array}{l}
\displaystyle
\exp (B_g t/2) ,
\\[2ex]
\displaystyle 
(1 - t/m_g^2)^{-2} ,
\end{array}
\right.
\label{twogl_exp_dip}
\ee
where the parameters $B_g$ and $m_g$ are functions of $x$ and $Q^2$.
The two parametrizations give very similar results if the functions 
are matched at $|t| = 0.5 \, \text{GeV}^2$, where they are best 
constrained by present data (see Fig.~3 of Ref.~\cite{Frankfurt:2006jp});
this corresponds to
\be
B_g \;\; = \;\; 3.24/m_g^2 .
\label{dip_exp}
\ee
We use both parametrizations in our studies below; 
the difference between the results serves as
an estimate of the uncertainty due to our lack of precise knowledge
of the shape. The corresponding spatial distributions of gluons in 
the transverse plane, Eq.~(\ref{rhoprof_def}), are given by
\be
F_g (x, \rho | Q^2) \;\; = \;\; 
\left\{ \begin{array}{l}
\displaystyle 
(2 \pi B_g)^{-1} \, \exp [-\rho^2 / (2 B_g)] ,
\\[2ex]
\displaystyle 
[m_g^2/(2\pi)] \; (m_g \rho/2) \; K_1 (m_{g} \rho ) ,
\end{array}
\right.
\label{f_rho_param}
\ee
where $K_1$ denotes the modified Bessel function.

Most of the experimental information on the nucleon's two--gluon
form factor comes from $J/\psi$ photoproduction, which probes 
the gluon GPD at an effective scale $Q^2 \approx \, 3 \, \text{GeV}^2$, 
determined by the average transverse size of the $c\bar c$
pair during its interaction with the target, and momentum
fractions of the order $x \sim M_{J/\psi}^2/W^2$ \cite{Frankfurt:1997fj}. 
When extracting the two--gluon form factor from the slope of the $J/\psi$
differential cross section, a correction is made for the effect
of the finite $J/\psi$ size on the observed $t$--dependence,
\be
B_{J/\psi} \;\; = \;\; B_g + \Delta B ,
\label{bpsi_bg_corr}
\ee
where $\Delta B \approx 0.3 \, \text{GeV}^2$ from a dipole model
estimate \cite{Frankfurt:1997fj}. 

\begin{figure}
\includegraphics[width=.48\textwidth]{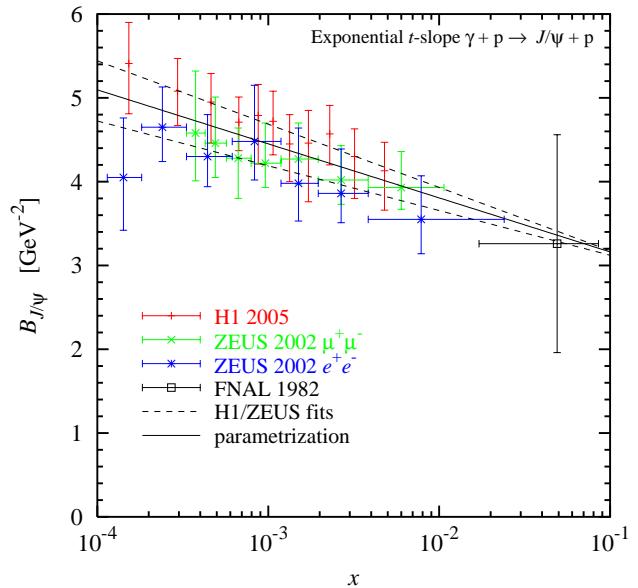}
\caption[]{The exponential $t$--slope, $B_{J/\psi}$, of the differential
cross section of exclusive $J/\psi$ photoproduction measured in the 
FNAL E401/E458 \cite{Binkley:1981kv}, HERA H1 \cite{Aktas:2005xu}, 
and ZEUS \cite{Chekanov:2002xi} experiments, as a function of 
$x = M_{J/\psi}^2/W^2$. (In the H1 and ZEUS results the quoted
statistical and systematic uncertainties were added linearly.)
The dashed lines represent the published two--dimensional fits to the 
H1 and ZEUS data \cite{Aktas:2005xu,Chekanov:2002xi}. 
The parameter $B_g$ in the exponential two--gluon form factor 
Eq.~(\ref{twogl_exp_dip}) is related to the measured $J/\psi$ 
slope by Eq.~(\ref{bpsi_bg_corr}). Our parametrization 
Eqs.~(\ref{bg_param})--(\ref{bg_param_last}) is shown by the solid line.} 
\label{fig:bpsi}
\end{figure}

The data from the FNAL E401/E458 broadband beam 
experiment at $\langle x \rangle = 0.05$ \cite{Binkley:1981kv}, 
in which the recoiling proton was detected, are described
by an exponential two--gluon form factor with
$B_g = 3.0 \, \text{GeV}^{-2}$ (see Fig.~\ref{fig:bpsi}),
albeit with large errors, or a corresponding dipole form factor 
with $m_g^2 = 1.1 \, \text{GeV}^2$. Comparison with the mass parameter 
in the dipole parametrization of the nucleon's electromagnetic (Dirac) 
form factor, $m_{\rm em}^2 = 0.7\, \text{GeV}^2$, indicates that at 
these values of $x$ the average transverse gluonic radius squared 
is only $\sim 0.6$ of the transverse electromagnetic radius squared. 
%
%

The HERA data at $x < 10^{-2}$ \cite{Aktas:2005xu,Chekanov:2002xi} 
indicate that the transverse gluonic size of the nucleon increases 
with decreasing $x$ (see Fig.~\ref{fig:bpsi}). 
In the region $x \sim \textrm{few} 10^{-2}$ this effect is partly 
due to the contribution of the nucleon's pion cloud, which is 
strongly suppressed at $x > 0.1$ and fully developed only for 
$x < 0.01$ \cite{Strikman:2003gz}. Another contribution may arise
from Gribov diffusion, which is suppressed by the hard scale
$Q^2$ but still not negligible. Over the range covered by HERA,
the rate of increase of the gluonic size can be parametrized by an 
effective Regge slope, $\alpha'_g$, which is smaller than that for 
corresponding soft processes, $\alpha'_{\rm soft} = 0.25 \, \text{GeV}^{-2}$. 
Averaging the fits to the HERA H1 and ZEUS 
data \cite{Aktas:2005xu,Chekanov:2002xi}, 
and accounting for the correction Eq.~(\ref{bpsi_bg_corr}),
we parametrize the $x$--dependence of the gluonic exponential 
slope as (here $Q^2 = 3 \, \text{GeV}^2$)
\bea
B_g (x) &=& B_{g0} \; + \; 2 \alpha'_g \; \ln (x_0/x) , 
\label{bg_param}
\\
x_0 &=& 0.0012, \\
B_{g0} &=& 4.1 \; ({}^{+0.3}_{-0.5}) \; \text{GeV}^{-2}, 
\\
\alpha'_g &=& 0.140 \; ({}^{+0.08}_{-0.08}) \; \text{GeV}^{-2}. 
\label{bg_param_last}
\eea
The uncertainties in parentheses represent a rough estimate
based on the range of values spanned by the H1 and ZEUS fits, 
with statistical and systematic uncertainties added linearly.
One sees from Fig.~\ref{fig:bpsi} that the fit to the HERA data 
consistently extrapolates to the FNAL data point. The corresponding
dipole parametrization obtained via Eq.~(\ref{dip_exp}) is
close to the one used in our previous study \cite{Frankfurt:2003td}.

The transverse spatial distribution of partons also changes with 
the resolution scale, $Q^2$, as a result of DGLAP evolution. 
Generally, the partons observed at a given momentum fraction $x$ 
and scale $Q^2$ are decay products of partons with $x' > x$ which
existed at a lower scale, $Q_0^2$. In the leading--twist 
approximation the decay happens locally in transverse space.
As a result, the transverse size observed at fixed $x$ shrinks
with increasing $Q^2$, because the decaying partons at the lower
scale had larger momentum fractions 
and were localized in a smaller transverse area. 
In order to calculate the change of the transverse spatial
distribution of gluons with $Q^2$ one would need to know the 
spatial distributions of both gluons and singlet quarks at the 
input scale for all $x' > x$, where they are only poorly 
constrained by present data. Numerical studies based on a
simple parametrization \cite{Frankfurt:2003td} suggest that
the evolution effect is small for $Q^2 > 3 \, \text{GeV}^2$.
The average transverse size $\langle \rho^2 \rangle_g$ at
$x \sim 10^{-3}$ decreases by $\sim 15\%$ between $Q^2 = 3$ and
$10^4 \, \textrm{GeV}^2$, while the effective value of $\alpha'_g$
in this $x$--region drops by about half. 
We note that the $J/\psi$ electroproduction data from 
HERA \cite{Chekanov:2004mw,Aktas:2005xu}
provide some indication that the effective $\alpha'_g$ may be 
smaller than in photoproduction, although the results are not
fully conclusive. In any case, the $\alpha'_g$
in the parametrization Eqs.~(\ref{bg_param})--(\ref{bg_param_last}) 
can be considered
as an upper limit at values of $Q^2 \gtrsim 10\, \textrm{GeV}^2$,
as are of interest in the applications here.

Comparatively little is known about the transverse distribution of
singlet quarks ($q + \bar q$) at small $x$. Comparison of the HERA 
deeply--virtual Compton scattering \cite{Aaron:2007cz}
and $J/\psi$ production data indicates 
that singlet quarks at $x < 10^{-2}$ are distributed over a larger 
transverse area than the gluons, in qualitative agreement with 
theoretical arguments based on the pion cloud contribution to the
parton densities at large $b$ \cite{Strikman:2009bd}.
In the applications here we are concerned with gluon--induced 
processes; parametrizations similar to Eq.~(\ref{twogl_exp_dip})
could be formulated also for the quark distributions.
\section{Impact parameter distribution of proton--proton collisions}
\label{sec:impact}
\begin{figure}
\includegraphics[width=.22\textwidth]{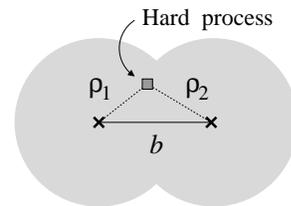}
\caption[]{Overlap integral of the transverse spatial parton
distributions, defining the impact parameter distribution of
$pp$ collisions with a hard parton--parton process, Eq.~(\ref{P_2_def}).}
\label{fig:overlap}
\end{figure}

Using the information on the transverse spatial distribution 
of partons in the nucleon, one can infer the distribution of
impact parameters in $pp$ collisions with hard parton--parton
processes. While not directly observable, the latter
determines the spectator interactions in such collisions and
thus can be studied indirectly through measurements of the
correlation of hard processes with final--state properties.
The hard parton--parton process is effectively pointlike in 
transverse space compared to the typical scale of variation of 
the transverse distributions of partons in the colliding hadrons.
The impact parameter distribution of $pp$ events
with a hard gluon--gluon process is thus given by the normalized
geometric probability for two gluons to collide at the same
point in transverse space:
\bea
P_2 (x_1, x_2, b|Q^2) &\equiv&
\int \! d^2\rho_1 \int \! d^2\rho_2 \; 
\delta^{(2)} (\bm{b} - \bm{\rho}_1 + \bm{\rho}_2 )
\nonumber \\
&\times& F_g (x_1, \rho_1 |Q^2 ) \; F_g (x_2, \rho_2 |Q^2) ,
\label{P_2_def}
\eea
where $b \equiv |\bm{b}|$ is the $pp$ impact parameter and 
$\rho_{1, 2} \equiv |\bm{\rho}_{1,2}|$ the transverse distances
of the two gluons from the center of their parent protons
(see Fig.~\ref{fig:overlap}) \cite{Frankfurt:2003td}. 
It satisfies the normalization condition 
$\int d^2b \, P_2 (x_1, x_2, b |Q^2) = 1$. 
With the parametrizations of Eq.~(\ref{twogl_exp_dip}) 
the convolution integral in Eq.~(\ref{P_2_def}) 
can easily be evaluated analytically. In the case of symmetric
collisions ($x\equiv x_1 = x_2$) we find
\be
P_2 (x, b| Q^2) \; = \; \left\{
\begin{array}{l}
\displaystyle
(4\pi B_g)^{-1} \, \exp [-b^2/(4 B_g)] ,
\\[2ex]
\displaystyle
[m_g^2 /(12\pi)] \, (m_g b/2)^3 \, K_3 (m_{g} b) ,
\label{P_2_exp_dip}
\end{array}
\right.
\ee 
where the parameters $B_g$ and $m_g$ are taken at the appropriate
values of $x$ and $Q^2$.
%
%

The impact parameter distribution in minimum--bias inelastic $pp$
collisions can be inferred from the $pp$ elastic scattering amplitude,
which incorporates the information on the $pp$ total cross section
through the unitarity relation. It is given by
\be
P_{\text{in}} (s, b) \;\; = \;\; 
[ 1 - |1 - \Gamma (s, b)|^2]\,  / \sigma_{\text{in}}(s) ,
\label{P_in_def}
\ee
where $\Gamma (s, b)$ is the profile function of the $pp$ elastic
amplitude in the conventions of Ref.~\cite{Frankfurt:2003td}
and the inelastic cross section $\sigma_{\rm in} (s)$ is given 
by the integral $\int d^2 b$ of the expression in brackets, such that 
$\int d^2 b \, P_{\text{in}} (s, b) = 1$. For the purpose
of the present study we employ a simple analytic parametrization of 
the profile function which satisfies unitarity and reflects the 
approach to the black--disk regime ($\Gamma \rightarrow 1$) 
at small impact parameters:
\be
\Gamma (s, b) \;\; = \;\; \Gamma_0 \, 
\exp \{ -b^2/ [2 B(s)] \} ,
\label{Gamma_gaussian}
\ee
where $\Gamma_0 = 1$ and the slope parameter is given in terms
of the total cross section as $B(s) = \sigma_{\rm tot}(s)/(4\pi)$;
the inelastic cross section for this profile is
$\sigma_{\rm in}(s) = 3\pi B(s)$. For the total cross section we 
use the extrapolation suggested by the COMPETE 
Collaboration \cite{Cudell:2002xe}, which gives 
$B(s) = 20.2 \; (22.8) \; \text{GeV}^{-2}$ at 
$\surd s = 7 \; (14) \; \text{TeV}$. The uncertainty in the profile 
function at LHC energies is dominated by that of the total 
cross section. The impact parameter
distributions calculated with Eq.~(\ref{Gamma_gaussian}) provide
a fully satisfactory representation of those obtained with more 
elaborate parametrizations of the $pp$ elastic amplitude, 
see Fig.~1 of Ref.~\cite{Frankfurt:2006jp} and references therein.

Using the above expressions we can now study the influence of the
trigger conditions on the impact parameter distribution of $pp$ 
events at the current LHC energy $\surd s = 7 \, \textrm{TeV}$.
The present experiments typically consider a jet trigger near 
zero rapidity, $y_1 \approx 0$, and study the characteristics of the 
underlying events as a function of the transverse momentum $p_T$
of the highest--momentum particle in the pseudorapidity interval 
$-2.5 < \eta < 2.5$. In this setting one integrates over the
energy of the balancing jet (as well as that 
of other jets which might arise from higher--order processes),
which effectively amounts to integrating over the momentum
fraction of the second parton, $x_2$, at fixed $x_1$. Since the 
distribution is symmetric in the rapidity of the balancing jet, $y_2$, 
and the variation of the transverse distribution of partons 
with $x$ is small, \textit{cf.}\ 
Eqs.~(\ref{bg_param})--(\ref{bg_param_last}) and Fig.~\ref{fig:bpsi}, 
we can to a good approximation set $y_2 = 0$ and thus
take $x_{1, 2}$ at the average point
\be
x_1 \; = \; x_2 \; = \; 2 p_T/\sqrt{s} .
\ee
%
%
\begin{figure}
\includegraphics[width=0.48\textwidth]{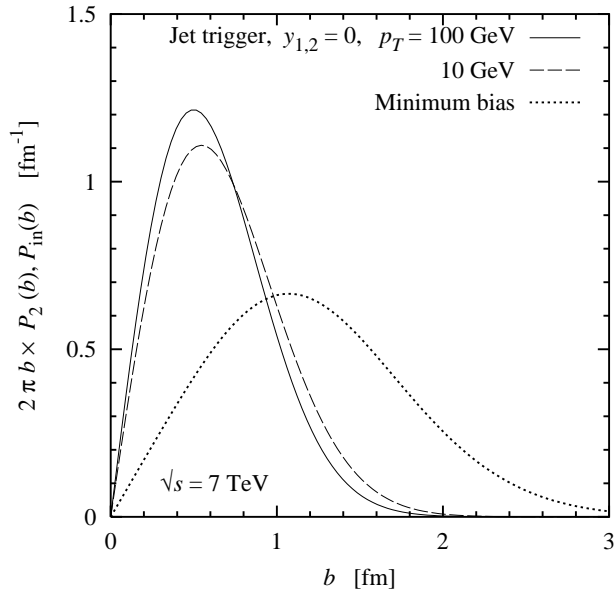}
\caption[]{Impact parameter distributions of inelastic $pp$ 
collisions at $\surd s = 7 \, \text{TeV}$.
\textit{Solid (dashed) line:} Distribution of events with a 
dijet trigger at zero rapidity, $y_{1, 2} = 0$, \textit{cf.}\ 
Eq.~(\ref{P_2_exp_dip}), for $p_T = 100 \, (10) \, \text{GeV}$ . 
\textit{Dotted line:} Distribution of minimum--bias inelastic events, 
\textit{cf.}\ Eq.~(\ref{P_in_def}).}
\label{fig:pb}
\end{figure}
The scale at which the parton densities are probed is of the order 
$Q^2 \sim p_T^2$, with a coefficient which remains
undetermined at leading--order accuracy. Generally, we expect
the impact parameter distribution in events with such a jet trigger
to become narrower with increasing $p_T$, because the transverse
distribution of partons shrinks both with increasing $x_{1,2}$ and with
increasing $Q^2$. The impact parameter distributions 
with a jet trigger of $p_T = 10$ and $100 \, \textrm{GeV}$
are presented in Fig.~\ref{fig:pb}. Shown are the results obtained
with the exponential parametrization of Eq.~(\ref{P_2_exp_dip})
and Eqs.~(\ref{bg_param})--(\ref{bg_param_last}); 
the dipole form leads to comparable results.
One sees that the change of the width of this distribution with $p_T$ 
is rather small, because the transverse distribution of gluons changes 
only little with $x$ in the range explored here; account of the 
$Q^2$ dependence of the transverse distribution of gluons 
would lead to an additional small change. One also sees that the
impact parameter distributions with the jet trigger are much
narrower than that in minimum--bias inelastic events at the same energy. 
This quantifies the two--scale picture of transverse nucleon structure
summarized in Fig.~\ref{fig:percent}.

%
%
\begin{figure}
\includegraphics[width=.48\textwidth]{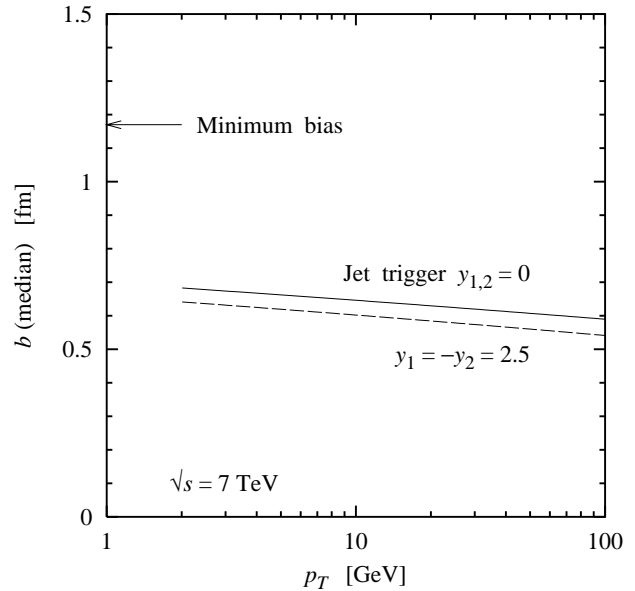}
\caption[]{Median impact parameter $b\textrm{(median)}$
of events with a dijet trigger, as a function of the transverse 
momentum $p_T$, \textit{cf.}\ Fig.~\ref{fig:pb}. 
\textit{Solid line:}  Dijet at zero rapidity $y_{1,2} = 0$.
\textit{Dashed line:} Dijet with rapidities $y_{1,2} = \pm 2.5$. 
The arrow indicates the median $b$ for minimum--bias inelastic events.}
\label{fig:medp2}
\end{figure}
The median impact parameter in dijet events, 
defined as the value of $b$ for which
the integral over $P_2$ reaches the value 1/2, is shown 
in Fig.~\ref{fig:medp2} as a function of $p_T$. For the parametrizations 
of Eq.~(\ref{P_2_exp_dip}) it is given by $b \text{(median)} = 
1.67 \, \surd B_g$ and $3.08 \, m_g^{-1}$, respectively.
The results obtained with the exponential and dipole form factors 
differ only by a few percent
if the parameters are related by Eq.~(\ref{dip_exp}),
indicating that the uncertainty resulting from our imperfect knowledge 
of the shape of the transverse spatial distribution of gluons 
is small. The uncertainty in $b \text{(median)}$ resulting from the 
uncertainty of $B_{g0}$ in the parametrization 
Eqs.~(\ref{bg_param})--(\ref{bg_param_last}) 
is less than $\pm 10\%$ at $p_T \sim \textrm{few GeV}$.
It is seen that the median $b$ in jet events
drops only very weakly as a function of $p_T$ for all values
above $\sim 2 \, \textrm{GeV}$. 
We estimate that account of the $Q^2$ dependence of 
the transverse distributions due to DGLAP evolution would change the 
results in Fig.~\ref{fig:medp2} by less than $\sim 5 \%$.
Also shown is the median $b$ with a trigger on a jets at non-zero 
rapidity $y_1 = - y_2 = 2.5$, which amounts to an effective increase 
of $x_{1, 2}$ by a factor $\cosh y \approx 6$, \textit{cf.}\ 
Eq.~(\ref{x_1_2_y}) and the discussion in Sec.~\ref{sec:rapidity}.
In all cases, the median impact parameter 
in jet events is far smaller than that in minimum--bias collisions, 
which is given by $b\text{(median)} = 1.32 \, \surd B$ for the 
parametrization of Eq.~(\ref{Gamma_gaussian}).

To conclude this discussion, a comment is in order concerning the 
interpretation of the impact parameter distributions in $pp$ events 
with hard processes. Our analysis based on Eq.~(\ref{P_2_def}) shows 
that $pp$ events with at least one hard process (and no other requirements) 
are on average more central than minimum--bias inelastic events. 
This statement concerns the \textit{relative} distribution of impact 
parameters in a collective of inelastic $pp$ events and how it is changed 
by imposing the requirement of a hard process. One should not confuse 
this with statements about the \textit{absolute} probability
for a hard process (in a certain rapidity interval) in a $pp$
collision at certain impact parameters. In fact, the analysis
of Refs.~\cite{Rogers:2008ua,Rogers:2009ke} shows that there
can be a substantial absolute probability for a hard process
in $pp$ collisions at large $b$, and that unitarity places
non--trivial restrictions on the dynamics of hard 
interactions in peripheral collisions.
\section{Transverse multiplicity as an indicator of hard dynamics}
\label{sec:multiplicity}
The estimates of the previous section show that $pp$ events
with a hard parton--parton collision are much more central than 
minimum--bias events, and that the average impact parameters change 
only very little for $p_T$ above $\sim 2 \, \text{GeV}$.
At the same time, it is known that the overall event characteristics, 
such as the average multiplicity, depend strongly on the centrality 
of the underlying $pp$ collision. Combining these two observations,
we can devise a practical method to determine down to which values 
of $p_T$ mid--rapidity particle production is predominantly
due to hard parton--parton collisions. The observable of interest
is the transverse multiplicity, measured in the direction perpendicular
to the transverse momentum of the trigger particle or jet. It is not 
\textit{directly} affected by the multiplicity associated with the trigger 
or balancing jets, but is \textit{indirectly} correlated with the 
presence of a hard process because of its dependence on the centrality.

\begin{figure}
\includegraphics[width=0.48\textwidth]{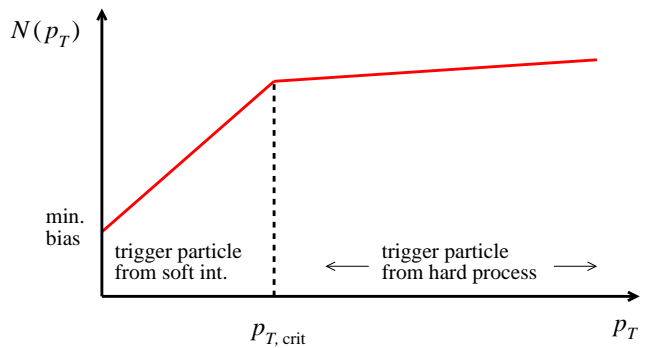}
\caption{Schematic illustration of the expected dependence of the
transverse multiplicity, $N(p_T)$, on the $p_T$ of the trigger.}
\label{fig:multpt}
\end{figure}

Based on the results of Figs.~\ref{fig:pb} and \ref{fig:medp2}
we predict that the transverse multiplicity should be practically
independent of $p_T$ of the trigger as long as the trigger particle
originates from a hard parton--parton collision which ``centers''
the $pp$ collision. Furthermore, the transverse multiplicity 
in such events should be significantly higher than in minimal--bias 
inelastic events, since the known mechanisms of particle production
--- minijet interactions, multiple soft interactions, \textit{etc.} ---
are much more effective in central collisions. When measuring the
transverse multiplicity as a function of $p_T$ of the trigger, 
we thus expect it to increase from its minimum--bias value at 
low $p_T$ and become approximately constant at $p_T \sim \textrm{few GeV}$
(see Fig.~\ref{fig:multpt}).
The point where the transition happens, $p_{T, {\rm crit}}$, 
indicates the critical value of $p_T$ above which particle production
is dominated by hard parton--parton processes.
%
%

Interestingly, the predicted increase and eventual flattening of the 
transverse multiplicity agrees well with the pattern observed in the
existing data. At $\surd s = 0.9\, \textrm{TeV}$ the transition occurs 
approximately at 
$p_{T, {\rm crit}} \approx 4\, \textrm{GeV}$ \cite{Khachatryan:2010pv},
at $\surd s = 1.8\, \textrm{TeV}$ at 
$p_{T, {\rm crit}} \approx 5\, \textrm{GeV}$ \cite{Affolder:2001xt},
and the preliminary data at $7\, \textrm{TeV}$ indicate 
somewhat larger values of $p_{T, {\rm crit}} = 6-8\, \textrm{GeV}$
\cite{EmilyNurse,Lucaroni}. We thus conclude that the minimum $p_T$ 
for hard particle production increases with the collision energy.
Note that we consider here an inclusive trigger; the procedure adopted 
in the experimental analysis (selection of the fastest particle 
in the measured rapidity interval) somewhat enhances the contribution 
of soft mechanisms in particle production. 
   
It is worth noting that the overall pattern described here is 
reproduced by the tunes of current MC models; \textit{cf.}\ the comparisons 
in Refs.~\cite{Affolder:2001xt,EmilyNurse,Khachatryan:2010pv,Lucaroni}. 
This is because these models effectively include the key feature used in 
our analysis --- the narrow impact parameter distribution of dijet events 
(although $\langle b^2 \rangle$ in these models is too small
by a factor $\sim 2$), and impose a cutoff on the minimal $p_T$ 
of the minijets. Our point here is that the observed pattern can
be explained naturally on the basis of the transverse geometry 
of $pp$ collisions with hard processes, without involving detailed 
models. This allows one to determine in a model--independent way 
where the dominant dynamics in particle production changes from soft 
interactions to hard parton--parton processes.

For $p_T$ lower than $p_{T, {\rm crit}}$ the relative contribution 
of hard processes to particle production starts to decrease.
In terms of the transverse geometry, this means that the observed 
trigger particle can, with some probability, originate from either 
peripheral or central collisions in the sense of Fig.~\ref{fig:percent}.
We can estimate the fraction of particles produced by hard 
interactions in this ``mixed'' region in a simple two--component
model, based on the observation that the effective impact parameters
in soft collisions are much larger than those in hard events
and do not change much with transverse momenta of the produced 
particles \footnote{This assumption 
is certainly not correct if the maximum $p_T$ of particles produced 
in a rapidity interval of a few units is taken too low. This would 
push the impact parameter to very large values and would likely enhance 
the contribution of double diffraction.}. Thus, we assume that: (\textit{i}) 
a trigger particle observed at given $p_T$ originated with a probability 
$\lambda_{\rm hard}(p_T)$ from a hard process, and with probability 
$1- \lambda_{\rm hard}(p_T)$ from soft interactions; (\textit{ii})
the average impact parameters in both classes of collisions do not 
depend of the $p_T$ of the trigger. This allows us to write the
$p_T$ dependence of the transverse multiplicity in the form
\be
N(p_T) \;\; = \;\; \lambda_{\rm hard}(p_T) N_{\rm hard} 
+ [1 - \lambda_{\rm hard}(p_T)]  N_{\rm soft} ,
\label{N_mixed}
\ee
where $N_{\rm hard}$ and $N_{\rm soft}$ are independent of $p_T$.
Assuming that for some sufficiently small $p_T$ cutoff 
$\lambda_{\rm hard}(p_T)$ is close to zero, we can determine 
$N_{\rm soft}$, which corresponds to the minimum--bias impact
parameter distribution presented in Fig.\ref{fig:pb}, and
use it to determine $\lambda_{\rm hard} (p_T)$ for $p_T$ smaller than
$p_{T, {\rm crit}}(s)$ via Eq.~(\ref{N_mixed}). The data indeed indicate 
that $N_{\rm hard} \gg N_{\rm soft}$; so in the region of $p_T$ where 
$N(p_T) /N_{\rm hard} \ge 1/3$ our estimate is not sensitive
to the exact value of $N_{\rm soft}$. By inspection of the data 
we conclude that the contribution of the hard mechanism drops to 
about half of the total yield for 
$p_T \approx 1.5-2, 2-2.5, 3-4 \, \textrm{GeV}$ for
$\surd s = 0.9, 1.8, 7 \, \textrm{TeV}$.
 
It is also of interest that for $p_T > p_{T, {\rm crit}}$ the transverse
multiplicity appears to increase with the collision energy faster 
than the average multiplicity \cite{EmilyNurse,Lucaroni}.
In the leading--twist approximation the perturbative contribution 
is proportional to the product of the gluon densities at small $x_{1, 2}$
and thus scales as $(\sqrt{s})^{2\lambda}$, where $\lambda$ is the 
exponent of the gluon density, $x g(x, Q^2) \propto x^{-\lambda}$, 
and takes on values $\lambda \sim 0.2-0.3$ in the $Q^2$ region of
interest here. This is roughly consistent with the factor $\sim 2$
increase of the observed transverse multiplicity between 
$\surd s = 0.9$ and 7 TeV, which suggests scaling as 
$(\sqrt{s})^{0.34}$ \footnote{In the case 
when no particles with $p_T > 2\, {\rm GeV}$ are produced in 
the measured rapidity interval the multiplicity does not
increase with $s$. Presumably this selection of events
corresponds to collisions at very large impact parameters, 
where soft interactions dominate and the change with energy is 
the slowest.}. We note that at very small values of $x_1$ or $x_2$ 
the leading--twist approximation breaks down because of the onset
of the black--disk regime in hard interactions, which generates
a new dynamical scale in the form of the gluon density per 
transverse area; see Ref.~\cite{Frankfurt:2003td} for
an estimate of the relevant values of $x_{1, 2}$ and $p_T$.
\section{Rapidity dependence as a test of universality}
\label{sec:rapidity}
The basic idea of our approach is that hard processes influence the event 
characteristics by selecting $pp$ collisions with small impact parameters.
Further interesting tests of these geometric correlations can be 
performed by measuring the dependence of event characteristics on
the rapidity of the jets in the trigger.

In production of dijets at non-zero rapidity only part of the 
center--of--mass energy of the colliding partons is converted to 
transverse energy, allowing one to probe larger momentum fractions 
$x_{1,2}$ at the same $p_T$. 
For jets with symmetric rapidities $y_1 = -y_2 \equiv y$,
\be
x_1 \; = \; x_2 \; = \; (2 p_T \cosh y) /\sqrt{s} .
\label{x_1_2_y}
\ee
Because partons with larger $x_{1,2}$ sit at smaller transverse
distances, the average impact parameters in $pp$ collisions
with a dijet trigger decrease with increasing $y$; however, 
the effect is small (see Fig.\ref{fig:medp2}). Observing the
approximate $y$--independence of the transverse multiplicity 
would test that the selection of central collisions does not
depend on the details of the hard process. Beyond that, we predict
a small increase in the transverse multiplicity if $y$ is increased 
away from zero at fixed $p_T$. In particular, such measurements could 
separate the effects of the $x$-- and $Q^2$--dependence of the 
transverse distribution of partons on the average impact parameters.
At lower $p_T$, the dependence of the transverse multiplicity on the 
rapidity would help to distinguish between the minijet and soft mechanisms 
of hadron production, as minijets are much more 
centered at small rapidities, while typical soft multi--ladder--type 
interactions lead to correlations over large rapidity intervals.
An additional advantage of measurements with the $y \neq 0$ trigger
is that the difference between the transverse, forward, and away--side 
regions of particles produced at mid--rapidity is much smaller
than for the $y = 0$ trigger. 

The selection of central $pp$ impact parameters by hard parton--parton 
processes could in principle be verified not only through the transverse 
multiplicity, but also by measuring event characteristics 
in \textit{rapidity regions} which are not directly affected by the
jet fragmentation of the partons in the trigger process. The identification 
of the jet fragmentation regions requires detailed modeling beyond 
the scope of the present investigation. Assuming that one could
reliably remove the fragmentation regions, several types of
interesting correlation measurements become possible. First,
we predict that in the remaining rapidity region the multiplicity 
should become, on average, isotropic in the transverse direction,
\textit{i.e.}, it should attain the value of the previously considered 
transverse multiplicity in all directions and be substantially higher 
than in minimum bias events. Second, this multiplicity should not
depend on the rapidity of the trigger, $y$, if $p_T > p_{T, {\rm crit}}$. 
Both measurements would directly attest to the universality of particle 
production in central $pp$ collisions.

The present LHC experiments use the central detectors to study 
the underlying event structure in the production of high--$p_T$ particles 
in the pseudorapidity interval $\eta = \pm 2.5$ (corresponding roughly 
to the same range in rapidity proper, $y$); the measurements may be
extended to $-5 \le \eta \le 5$ (CMS) and $-4.9 \le \eta \le 4.9$
(ATLAS) using forward detectors.
In production of two jets at $y_1 = -y_2 = y \approx 2$, assuming a rapidity 
interval of approximately $\pm 0.7$ for the fragmentation of either jet,
the rapidity region $\pm 1$ should be free of direct jet fragments
and could be used for the envisaged multiplicity measurements. 
Alternatively, one may consider a pair of jets at the same positive
rapidity, $y_1 = y_2 \equiv \bar y > 0$, and study the multiplicity in the 
negative rapidity region as a function of $\bar y$. 
The latter choice would have the advantage that the parton momentum 
fractions $x_{1, 2}$ change in different directions when increasing
$\bar y$ from zero, compensating the effect on the width of the
impact parameter distribution to first order.

An interesting phenomenon should occur when extending such measurements
with symmetric jets at $y_1 = -y_2 = y$ to larger rapidities. As discussed 
in Sec.~\ref{sec:parton}, the transverse size of the parton distribution 
decreases with increasing $x$. This leads to a seemingly paradoxical 
prediction, that the larger the rapidity interval $y_1 - y_2 = 2y$ 
between the jets, the larger the multiplicity in the mid--rapidity region. 
In other words, one expects long--range correlations in rapidity 
which are becoming stronger with increase of the rapidity interval. 
However, the effect is rather small; at $p_T = 5 \, \textrm{GeV}$ 
the median $pp$ impact parameter changes from 
$0.66 \, \textrm{fm}$ at $y = 0$ to $6.1 \, (0.55) \, \textrm{fm}$
at $y = 2.5 \, (5)$. Therefore the ability to measure over a wide
range of pseudorapidities $|\eta| < 5$ would be very helpful for
studying this effect.

Deviations from the predicted universality can arise as a result of
spatial correlations between partons involved in the hard collisions 
and those participating in spectator interactions \cite{Frankfurt:2004kn}.
At the relatively small $x$--values probed with the central detectors 
at LHC ($x < 10^{-2}$), such correlations are likely to depend weakly 
on $x$ and would not significantly affect the rapidity dependence.
In principle, the study of these deviations from universality may provide 
a new window on correlations in the nucleon's partonic wave function.
\section{Suggestions for future measurements}
\label{sec:future}
In addition to the studies described in Secs.~\ref{sec:multiplicity}
and \ref{sec:rapidity}, several other kinds of measurements could 
further explore the proposed connection between hard processes and 
the transverse multiplicity, or use it to investigate interesting 
aspects of QCD and nucleon structure.

\textit{Energy dependence of transverse multiplicity.} 
Measurement of the energy dependence of the transverse multiplicity
in jet events would, in effect, reveal the energy dependence of
the average multiplicity in central $pp$ collisions, which is of 
interest beyond the specific applications considered here.
In order to avoid change of the average impact parameters due to 
the $x$--dependence of the transverse distribution of partons, 
it would be desirable to compare data at the same values of $x_{1,2}$. 
When increasing the collision energy from $\surd s_0$ to $\surd s$ 
one thus needs to trigger on jets with rapidities scaled 
by $(1/2)\ln(s/s_0)$.

\textit{Double dijet trigger.} Further reduction of the effective
impact parameters in $pp$ collisions can be achieved with a trigger
on multiple hard processes \cite{Frankfurt:2003td}. In the absence
of transverse correlation between partons, the effective 
impact parameter distribution of events with two dijets would be
given by the (properly normalized) product of distributions 
$P_2$ in Eq.~(\ref{P_2_def}). In the simplest case of two dijets
with $x_1 = x_2 = x_3 = x_4$, the median $b$ in such events 
would be $1.18 \, \surd B_g \; (1.97 \, m_g^{-1})$, to be compared to
$1.67 \, \surd B_g \;(3.08 \, m_g^{-1})$ for single jet events.
Account of transverse correlations between partons reduces the
difference in effective impact parameters by about 
half \cite{Frankfurt:2003td}. In all, we expect a $15-20\, \%$
reduction of the median $b$ with a double dijet trigger, which
should manifest itself in a further increase of the transverse
multiplicity compared to single dijet events.

\textit{Other centrality triggers.} Knowledge of the dependence of 
the transverse multiplicity on $b$ would allow one to calibrate 
other triggers on central $pp$ collisions. In particular, it would be 
interesting explore triggers related to particle production
in the nucleon fragmentation regions, for example leading neutrons. 
Ultimately one aims here at designing a trigger on ultra--central
$pp$ collisions, in which the effective gluon densities would be 
comparable to those reached in heavy--ion collisions \cite{Drescher:2008zz}.

\textit{Quark vs.\ gluon--induced jets.} It would be interesting to 
compare the transverse multiplicities and other underlying event 
characteristics for quark--antiquark induced hard processes like
$W^{\pm}, Z^0$ production and gluon--gluon induced processes. 
Another possibility would be to consider large--$|\eta|$ dijet 
production and separate quark-- and gluon induced jets using the
different jet shapes.
This would allow one to observe a possible difference between the
transverse distributions of quarks and gluons, complementing
studies of hard exclusive processes in $ep/\gamma p$ scattering.

\textit{Deconvolution of impact parameter dependence.}
If the fluctuations of the multiplicity between events at a given 
impact parameter are not too large, one can attempt a program of 
deconvolution of the impact parameter dependence of the multiplicity, 
using information on the impact parameter dependence of dijet 
and 4--jet rates \cite{Rogers:2009ke}.
\section{Summary and discussion}
\label{sec:summary}
The transverse multiplicity in $pp$ collisions with a jet trigger 
provides an effective way to determine the rate of processes initiated 
by hard parton--parton interactions down to rather small transverse momenta.
Further analysis of the data can then establish whether
the observed particle production rates are consistent with perturbative 
QCD predictions. Studies of the dependence of the transverse multiplicity 
on the transverse momenta and rapidities of the jets, and on the collision 
energy, can provide a better understanding of the impact parameter 
dependence of the underlying event characteristics and allow one
to refine the use of the transverse multiplicity as an indicator
of the dynamics in particle production.

The results of the present study have implications also for the use
of hard processes in new particle searches. In the production of
a Higgs boson with a mass $M_H \sim 100 \, \text{GeV}$ the parton 
momentum fractions $x_{1, 2}$ are the same as in dijet events
at zero rapidity with $p_T = M_H/2 = 50 \, \text{GeV}$. The two
types of events thus involve the same average $pp$ impact parameters.
This allows one to describe the background to new particle production
processes much more accurately than on the basis of minimum--bias
event characteristics.

Our analysis relies crucially on information about the transverse
spatial distribution of gluons from exclusive $J/\psi$ 
photo/electroproduction and similar processes. While the region 
$x < 0.01$ has been covered by HERA, no data of comparable 
precision are available at larger $x$ (see Sec.~\ref{sec:parton}). 
This is unfortunate,
as the production of new particle with masses $\sim 10^2 \, \text{GeV}$
requires partonic collisions at precisely such momentum fractions.
The region $x > 0.01$ is also particularly interesting for nucleon 
structure studies \cite{Strikman:2003gz,Strikman:2009bd}. 
The $J/\psi$ results expected from the COMPASS experiment, 
as well as measurements with a future Electron--Ion Collider (EIC), 
thus would have a major impact on both areas of study.
\begin{acknowledgments}
We thank J.~D.~Bjorken for inspiring conversations on related aspects 
of $pp$ collisions, and P.~Bartalini, E.~Nurse, P.~Skands, and B.~Webber 
for very useful discussions of current MC models and the LHC data.
Two of us (LF and MS) would like to thank the Yukawa International
Program for Quark--Hadron Sciences for hospitality during 
the initial stage of this study. MS also acknowledges the hospitality 
of the CERN Theory Institute ``The first heavy--ion collisions at the LHC'' 
during part of the work on this project. 

This research was supported by the 
United States Department of Energy and the Binational Science Foundation. 
Notice: Authored by Jefferson Science Associates, LLC under U.S.\ DOE
Contract No.~DE-AC05-06OR23177. The U.S.\ Government retains a
non--exclusive, paid--up, irrevocable, world--wide license to publish or
reproduce this manuscript for U.S.\ Government purposes.
\end{acknowledgments}
\end{document}